# Suppression of spin-exciton state in hole overdoped iron-based superconductors

C. H. Lee[1], K. Kihou[1], J. T. Park[2], K. Horigane[3], K. Fujita[3], F. Waßer[4], N. Qureshi[4], Y. Sidis[5], J. Akimitsu[3], and M. Braden[4]

[1]*National Institute of Advanced Industrial Science and Technology (AIST), Tsukuba, Ibaraki 305-8568, Japan.* [2]*Heinz Maier-Leibnitz Zentrum (MLZ), Technische Universität München, D-85748 Garching, Germany,* [3]*Aoyama Gakuin University, Sagamihara 252-5258, Japan.* [4]*II. Physikalisches Institut, Universität zu Köln, 50937 Cologne, Germany,* [5]*Laboratoire Léon Brillouin (LLB), C.E.A./C.N.R.S., F-91191 Gif-sur-Yvette Cedex, France.* [*]*E-mail: c.lee@aist.go.jp*

**The mechanism of Cooper pair formation in iron-based superconductors remains a controversial topic. The main question is whether spin or orbital fluctuations are responsible for the pairing mechanism. To solve this problem, a crucial clue can be obtained by examining the remarkable enhancement of magnetic neutron scattering signals appearing in a superconducting phase. The enhancement is called spin resonance for a spin fluctuation model, in which their energy is restricted below twice the superconducting gap value ($2\Delta_s$), whereas larger energies are possible in other models such as an orbital fluctuation model. Here we report the doping dependence of low-energy magnetic excitation spectra in $Ba_{1-x}K_xFe_2As_2$ for $0.5<x<0.84$ studied by inelastic neutron scattering. We find that the behavior of the spin resonance dramatically changes from optimum to overdoped regions. Strong resonance peaks are observed clearly below $2\Delta_s$ in the optimum doping region, while they are absent in the overdoped region. Instead, there is a transfer of spectral weight from energies below $2\Delta_s$ to higher energies, peaking at values of $3\Delta_s$ for x = 0.84. These results suggest a reduced impact of magnetism on Cooper pair formation in the overdoped region.**

One of the most plausible mechanisms for Cooper pair formation in iron-based superconductors is spin-mediated superconductivity. In this model, the superconducting gap symmetry is predicted to be $s_{\pm}$ with opposite phases between electron and hole Fermi surfaces [1,2]. Another possibility is superconductivity induced by orbital fluctuations, where the gap symmetry of $s_{++}$ is expected with identical phases between electron and hole Fermi surfaces [3,4]. The orbital fluctuation model has been experimentally supported by ultrasound measurements [5], impurity effects of superconducting transition temperature ($T_c$) [6,7] and Raman scattering [8]. To judge which model is appropriate, further studies are required.

Direct insight into determining a superconducting mechanism can be obtained by the examination of spin resonance. In the spin fluctuation model, the resonance mode is interpreted as a bound spin-exciton state (particle-hole excitation) [1, 2, 9, 10]; therefore, its energy ($E_{res}$) must stay below $2\Delta_s$. In contrast, in the orbital fluctuation model, there is an enhancement of scattering at energies exceeding $2\Delta_s$ [11], and scattering at low energies is suppressed in the superconducting state. This can be explained by the suppression of particle-hole scattering at low energies due to the opening of the superconducting gap. Several inelastic neutron scattering (INS) experiments have previously been performed on spin resonance modes of FeAs-based superconductors, especially in electron-doped $BaFe_2As_2$ [12-23]. It has been found that the resonance correlates strongly with magnetism, supporting the spin-exciton model [13-16]. In contrast, for hole-doped $BaFe_2As_2$, doping dependence remains unclear in spite of higher $T_c$ [18-22].

Thermal conductivity measurements on $Ba_{1-x}K_xFe_2As_2$ suggest that the superconducting gap symmetry is transformed from s to d-wave with doping in a heavily overdoped region [24]. On the other hand, angle-resolved photoemission (ARPES) measurements demonstrate that the full s-wave gap varies to a nodal s-wave gap above x = 0.7 [25,26]. Although the gap symmetry in the heavily hole overdoped region remains controversial, there is a consensus that a superconducting gap structure changes dramatically with doping. To clarify how the relationship between magnetism and superconductivity varies with changing the gap structure, we studied the doping dependence of spin fluctuations as well as that of spin resonances in $Ba_{1-x}K_xFe_2As_2$ using the INS technique on single-crystalline samples.

Results

Low-energy spin fluctuations of nearly optimum hole-doped $Ba_{1-x}K_xFe_2As_2$ (x = 0.34) exhibit commensurate peaks at Q = (0.5, 0.5, $L$) [21]. These magnetic peaks split along the longitudinal direction forming incommensurate peaks at Q = (0.5 ± δ, 0.5 ± δ, $L$) for x > 0.5 (Fig. 1). Already for the x = 0.50, slightly overdoped sample with a $T_c$ of 36 K, double peaks were observed with an overlapping structure. These peaks separate completely as doping further increases, see also [19, 27]. Our results qualitatively agree with the previous INS study on powder samples [22] and demonstrate that the incommensurability δ increases with increasing doping level associated with the suppression of $T_c$ from the optimum-doped region. As shown in Fig. 1(b), the suppression of $T_c$ with increasing δ follows a parabolic behavior up to x = 0.66. This clear relationship suggests that the periodicity of the spin fluctuations has a considerable impact on superconductivity and that a commensurate structure is advantageous for achieving high $T_c$. The parabolic correlation, however, ends near x = 0.66, where $T_c$ drops dramatically. This suggests that the interplay between magnetism and superconductivity changes around x = 0.66.

Figure 2 and Supplementary Figure S2 show energy dependences of magnetic signals at T ~ $T_c$ and T < $T_c$ (See also Supplementary Fig. S1). Backgrounds were determined at the sides of the magnetic rods. Huge enhancement of signals below $T_c$ is observed at E = 15 meV for x = 0.50 samples, which can be identified as a spin resonance. The enhancement weakens with increasing doping, but still exhibits slight enhancement even in x = 0.84.

The imaginary part of dynamical magnetic susceptibility χ"(q,ω) was obtained by multiplying the net intensities by [1-exp($-\hbar\omega/k_B T$)] after normalizing them by acoustic phonon intensities and correcting for higher-order components in the incident beam monitor [Figure 3(a-d); Supplementary Fig. S3(a-d)]. We can obtain $\chi''(q,\omega)$ in semi-absolute units by this procedure, which allows us to quantitatively compare the magnetic signals within the series of $Ba_{1-x}K_xFe_2As_2$. $\chi''(q,\omega)$ at E ≤ 21meV in the normal state seems to be nearly independent of the hole concentration. Peak structures are observed around E = 12meV in all samples and the magnitude of $\chi''(q,\omega)$ is almost constant with doping. The absence of a sizeable variation of the normal-state magnetic response is remarkable in view of the pronounced changes of the Fermi surface topology reported for these compounds. In contrast, $\chi''(q,\omega)$ in the superconducting phases exhibits a dramatic doping dependence. At x = 0.50, well-defined spin resonance peak is observed at $E_{res}$ = 15 meV, where $\chi''(q,\omega)$ increases by a factor of four upon cooling [Fig. 3(a,e); Supplementary Fig. S3(a,e)]. On the other hand, signals below E = 10 meV are drastically suppressed below $T_c$, demonstrating the opening of a spin gap. As the hole doping is further increased, the resonance intensity decreases drastically in qualitative agreement again with the previous study on powders [22], while $E_{res}$ stays almost constant (Fig. 3; Supplementary Fig. S3). Finally, only slight differences are observed between $\chi''(q,\omega)$ at T = $T_c$ and at T << $T_c$ for x = 0.84 that cannot be attributed to a resonance mode. According to ARPES measurements, the value of the maximum $2\Delta_s$ is $7.5k_BT_c$ in the overdoped region [28], which corresponds to 23.3 meV for x = 0.50. Apparently, the obtained $E_{res}$ is well below $2\Delta_s = 7.5k_BT_c$ for x = 0.50. On the other hand, there is no comparable intensity enhancement below E = $7.5k_BT_c$ (7.1 meV for x = 0.84) under cooling for x = 0.84; instead, broad and slight enhancement was observed at higher energies near E = 10.5 meV, which is certainly above $7.5k_BT_c$. There is no evidence that a spin resonance mode appears below $2\Delta_s$ for x = 0.84. Assuming that the broad peak in the temperature difference around E = 10.5 meV can be associated with the resonance, $E_{res}$ exceeds $2\Delta_s$ above x = 0.77 [Fig. 3(i)].

The development of the spin resonances and the gap opening can also be confirmed by the temperature dependence of the scattering intensity. The intensity at E = $E_{res}$ increases and that at E << $E_{res}$ decreases below $T_c$ in the overdoped samples including x = 0.84 (Fig. 4). This ensures that the broad and slight intensity enhancement at x = 0.84 can be attributed to the appearance of superconductivity.

The doping dependences of the obtained magnetic parameters are summarized in

Fig. 5. Integrating $\chi''(q,\omega)/\omega$ over energies up to E = 18meV yields nearly doping independent values, suggesting that the related real part of the dynamical susceptibility defined as $\chi'(q,\omega) = 1/\pi \int \chi''(q,\omega)/\omega d\omega$ by the Kramers-Kronig relation has no direct relationship with $T_c$. Note that, however, the high-energy response may considerably decrease with the doping as it is suggested by the results for the end member $KFe_2As_2$ [29]. The doping dependent incommensurability indicates a clear relationship with $T_c$ up to x = 0.66 [Fig.1(b)]. In contrast, $\delta$ becomes insensitive to the doping level as well as to $T_c$ above x = 0.66, indicating an essential change of the coupling between spin fluctuations and superconductivity. Furthermore, the spin resonance modes vary drastically with doping. The dramatic suppression of the resonance intensities itself can be explained by the spin exciton model, where the resonance intensity is proportional to $2\Delta_s - E_{res}$ under the restriction that $E_{res}$ should be smaller than $2\Delta_s$. In fact, $E_{res}$ is smaller than $2\Delta_s$ and approaches $2\Delta_s$ with increasing doping, keeping $E_{res}/k_BT_c$ almost constant value of 5 below x = 0.66. On the other hand, $E_{res}/k_BT_c$ increases drastically above x = 0.66 and reaches values of 10.5 at x = 0.84. Such high value can no more be explained by the spin exciton model, which cannot yield a resonance feature above $2\Delta_s$ within the continuum of particle-hole excitations. Other mechanisms such as the orbital fluctuation model should be considered.

Discussion

In overdoped region, there is thus a dramatic change in the magnetic response in the superconducting state, which contrasts with the small degree of variation in the normal state. There can be various reasons for the suppression of the resonance mode. A decrease of the nesting properties can explain such suppression, but this should also have a strong impact on the normal state response. The complex gap structure in overdoped $Ba_{1-x}K_xFe_2As_2$ can effectively suppress the resonance signal, which arises from averaging particle-hole processes over the entire Fermi surface. Finally, the reduction of the correlation strength indicated by the reduced bandwidth of magnetic excitations in $KFe_2As_2$ [29] may also contribute to the suppression of the resonance; the reduced correlations prohibit a clear separation of the bound resonance state from the continuum of particle-hole excitations.

The intensity enhancement at $T < T_c$ above x = 0.77, on the other hand, cannot be explained by the spin-exciton model. One interpretation is a renormalization of the particle-hole continuum accompanied by the opening of the superconducting gap. This has been discussed in LiFeAs [23,30], where the suppression of the resonance mode resembles the present results in overdoped $Ba_{1-x}K_xFe_2As_2$. In LiFeAs, the crossover of the spectral weight lies still below the maximum values of $2\Delta_s$, like the present x = 0.77 results.

Another possibility is based on the orbital fluctuation model. This leads to the idea that spin fluctuation and orbital fluctuation models compete in iron-based superconductors. The spin fluctuation model can be dominant in the underdoped region, which is close to the three-dimensional AFM phase, as it is suggested by the strong resonance mode well below $2\Delta_s$. In fact, the dispersive and anisotropic characters of the spin resonance are observed in the underdoped and optimum doping regions, which can be well explained by the spin-exciton model [16]. In the overdoped region, on the other hand, the three-dimensional character of the AF correlations is completely lost and spin fluctuations appear at an incommensurate vector. Owing to the change of those magnetic circumstances, the relationship between spin fluctuation and superconductivity seems to vary dramatically in the overdoped region. A transition from spin fluctuation to orbital fluctuation or some other model can occur in overdoped region of $Ba_{1-x}K_xFe_2As_2$.

Methods

Single crystals of $Ba_{1-x}K_xFe_2As_2$ were grown by the self-flux method [31]. The obtained single crystals had a tabular shape with (001) planes as their surfaces. They were coaligned on a thin Al sample holder to increase their total volume for inelastic neutron scattering experiments. The total weights of single crystals were 1.32, 0.82, 0.69, 0.85 and 0.64 g for x = 0.50, 0.58, 0.66, 0.77 and 0.84, respectively. Composition was determined by the energy dispersive X-ray analysis as well as lattice constant c examined by x-ray diffraction. Composition distribution among each assembled single crystals was confirmed to be less than ± 0.04 in x value by examining lattice constant c of all single crystals using x-ray diffraction on both sides of sample surfaces. Superconducting transition temperatures were examined by SQUID magnetometer (Quantum Design MPMS) under a magnetic field of 10 Oe after zero-field-cooling. Onset temperatures of $T_c$ were 36, 30, 25, 16 and 11.5 K with a transition width of 6, 5, 8, 5, 4 K for x = 0.50, 0.58, 0.66, 0.77 and 0.84, respectively. Inelastic neutron scattering measurements were conducted at FRM ll on the triple-axis spectrometer PUMA and at LLB on the 2T1 spectrometer. The final neutron energy was fixed at $E_f$ = 14.7 meV by using double-focusing pyrolytic graphite crystals as a monochromator and analyzer. To remove higher-order neutrons, a pyrolytic graphite filter was inserted between the sample and the analyzer. Data of x = 0.50, 0.58, 0.66 and 0.77 were measured at PUMA and that of x = 0.84 was measured at PUMA and 2T1.

Acknowledgments

We would like to acknowledge discussions with K. Kuroki, H. Kontani and S. Onari. This work was supported by a Grant-in-Aid for Scientific Research B (No. 24340090) from the Japan Society for the Promotion of Science and by the Deutsche Forschungsgemeinschaft through the Priority Programme SPP1458 (BR2211/1-1).

Author contributions

C.H.L., J.T.P., K.H., F.W., N.Q., Y.S., and M.B. conducted the inelastic neutron scattering measurements and analyzed the data. K.K., K.H. and K.F. synthesized and characterized the single crystals. C.H.L., J.A, and M.B. designed and coordinated the experiment. All authors contributed to and discussed the manuscript.

Additional information

Competing financial interests: The authors declare no competing financial interests.

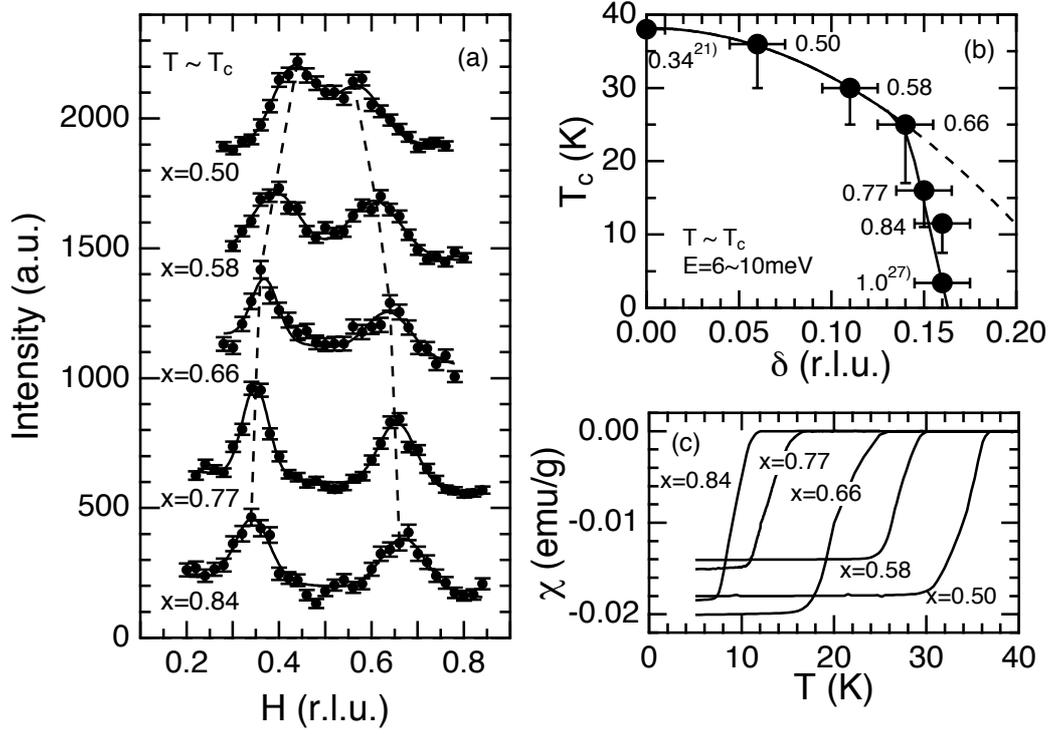

Figure 1. (a) Q-spectra of $Ba_{1-x}K_xFe_2As_2$ along $(H, H, 1)$ in the normal state. Energies and temperatures are E = 8 meV, T = 42 K for x = 0.50; E = 8 meV, T = 36 K for x = 0.58; E = 8 meV, T = 36K for x = 0.66; E = 8 meV, T = 22 K for x = 0.77 and E = 6 meV, T = 15 K for x = 0.84. (b) $T_c$ vs. incommensurability δ in the normal state close to $T_c$. Vertical error bars depict a superconducting transition width. Data for x = 0.34 and 1.0 are extracted from refs. 21 and 27, respectively. (c) Shielding signals measured under a magnetic field of H = 10 Oe.

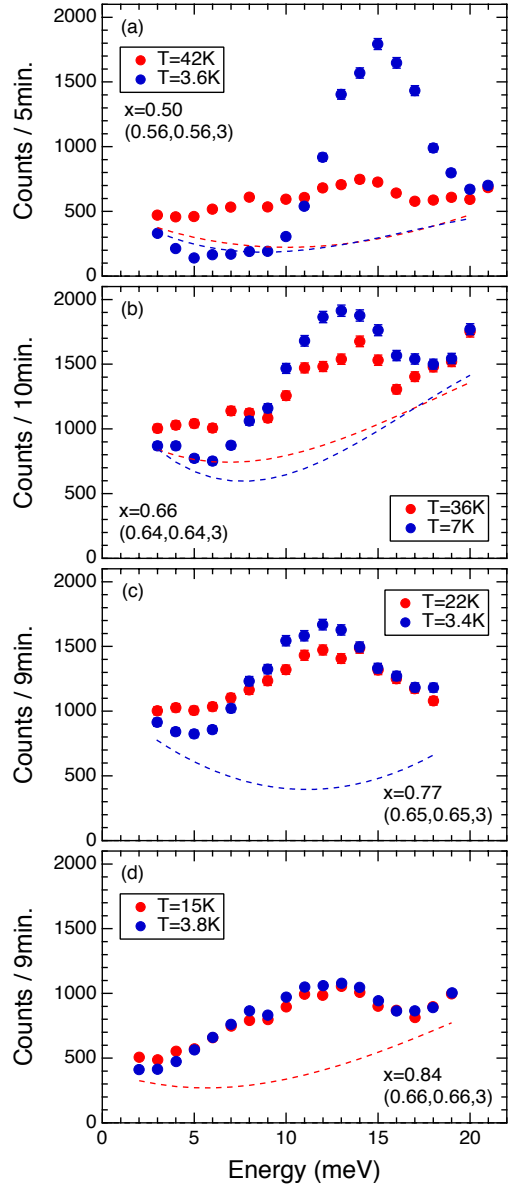

Figure 2. Energy dependences of magnetic signals at Q = (0.5 ± δ, 0.5 ± δ, 3) at T ~ $T_c$ and T < $T_c$ for (a) x = 0.50, (b) x = 0.66, (c) x = 0.77 and (d) x = 0.84. Dashed lines describe the background at T ~ $T_c$ (red) and T < $T_c$ (blue) determined by averaging the background at (a) (0.28,0.28,4), (0.72,0.72,3) and (0.695,0.695,0); (b) (0.48,0.48,3.65), (0.22,0.22,3) and (0.78,0.78,3); (c) (0.2,0.2,4.29) and (d) (0.5,0.5,3.69). Because $T_c$ is sufficiently low for x ≥ 0.77 samples, their backgrounds should be almost temperature-independent below $T_c$.

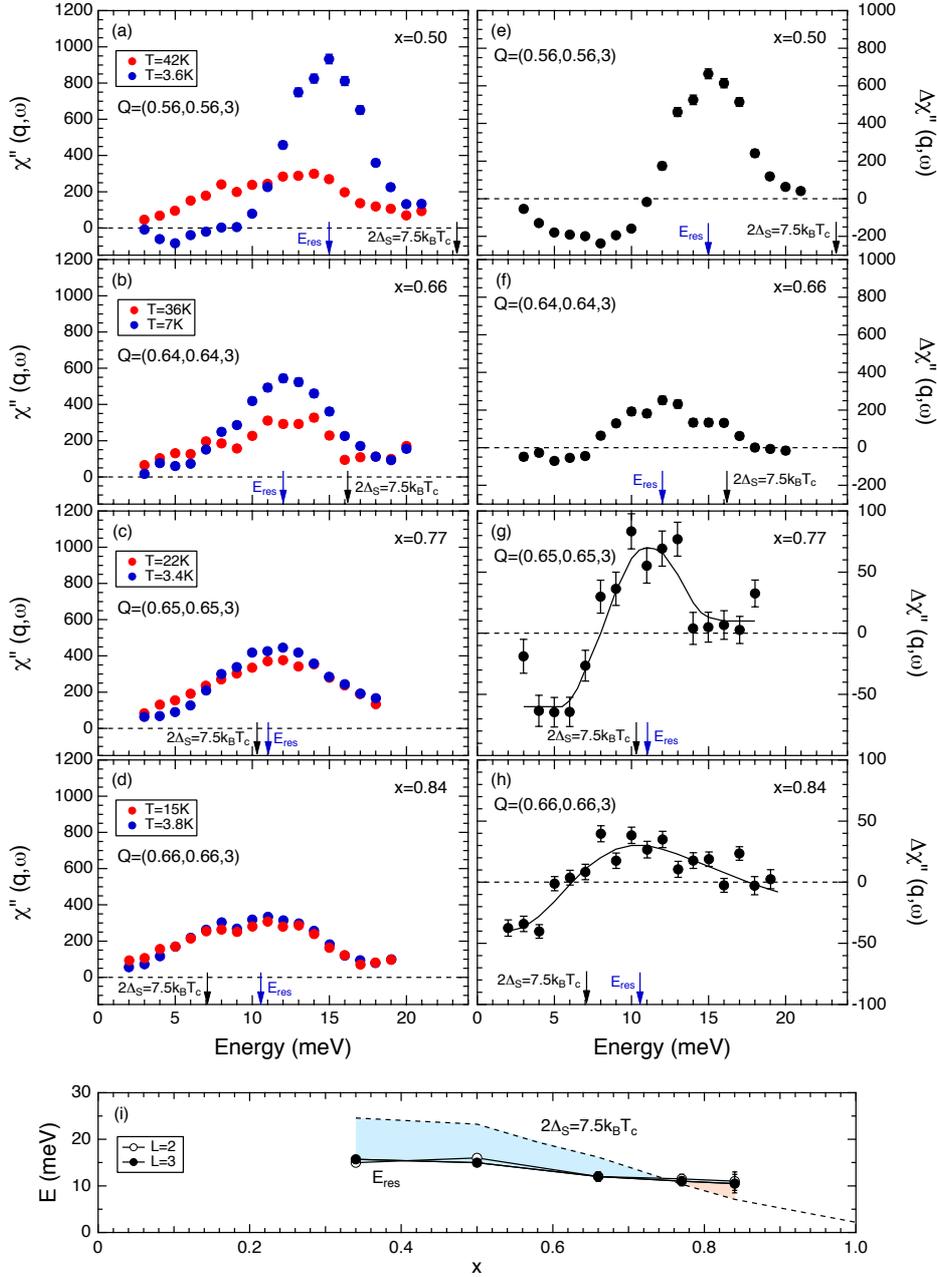

Figure 3. (a-d) Energy dependences of $\chi''(q,\omega)$ at peak positions in Q-scans at $T \sim T_c$ and $T < T_c$ for (a) $x = 0.50$, (b) $x = 0.66$, (c) $x = 0.77$ and (d) $x = 0.84$. Amplitudes were normalized by phonon scattering intensities. (e-h) Difference of $\chi''(q,\omega)$ between $T \sim T_c$ and $T < T_c$. (i) Doping dependences of $E_{res}$ at $L = 2$ and 3 with data of $x = 0.34$ extracted from ref. 21. $E_{res}$ is almost independent of $L$. The dashed line depicts the superconducting gap value of $2\Delta_s = 7.5k_BT_c$.

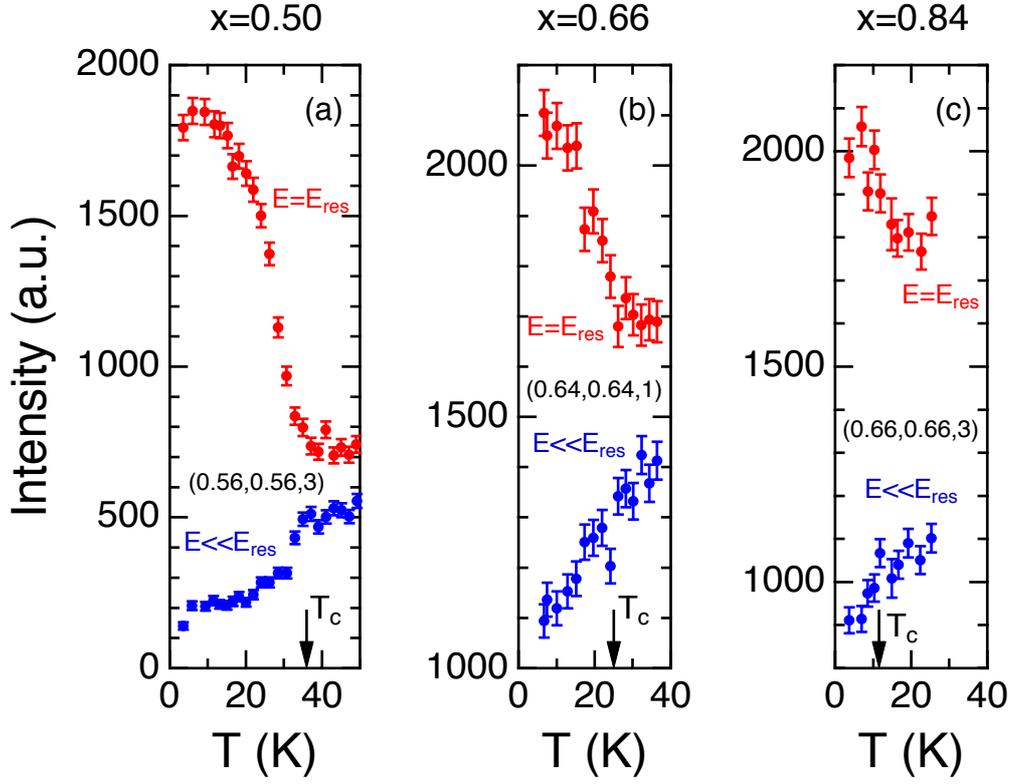

Figure 4. Temperature dependences of intensity at E = $E_{res}$ and E < $E_{res}$. (a) Q = (0.56, 0.56, 3), E = 15 meV and 5 meV for x = 0.50, (b) Q = (0.64, 0.64, 1), E = 12 meV and 3 meV for x = 0.66 and (c) Q = (0.66, 0.66, 3), E = 12 meV and 2 meV for x = 0.84. Intensities at E = $E_{res}$ increase with decreasing temperature below $T_c$, whereas those at E < $E_{res}$ decrease.

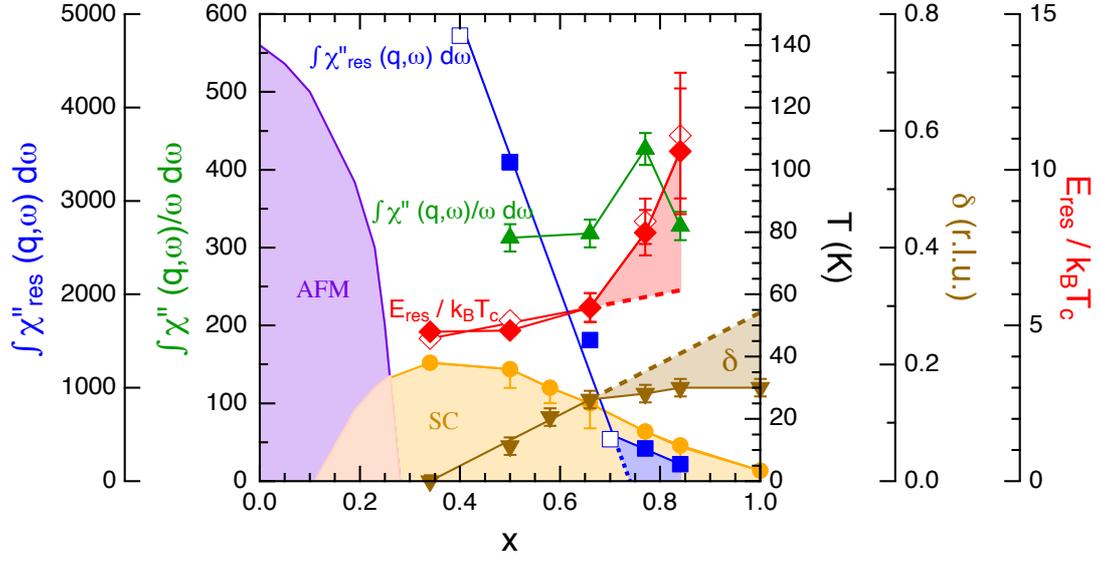

Figure 5. Phase diagram of $Ba_{1-x}K_xFe_2As_2$. Symbols denote $T_c$ (circle), $E_{res}/k_BT_c$ for $L$ = even with data of x = 0.34 extracted from ref. 21 (open diamond), $E_{res}/k_BT_c$ for $L$ = odd with data of x = 0.34 extracted from ref. 21 (closed diamond), energy-integrated $\Delta\chi''(q,\omega)$ around $E_{res}$ defined as $\int\chi''_{res}(q,\omega)d\omega$ (closed square), $\int\chi''_{res}(q,\omega)d\omega$ extracted from ref. 22 normalized at x = 0.50 (open square), energy integrated $\chi''(q,\omega)/\omega$ from 3 to 18 meV in a normal state (triangle) and incommensurability at T ~ $T_c$ (inverted triangle). Dashed lines are extrapolation.

# Suppression of spin-exciton state in hole overdoped iron-based superconductors

C. H. Lee[1], K. Kihou[1], J. T. Park[2], K. Horigane[3], K. Fujita[3], F. Waßer[4], N. Qureshi[4], Y. Sidis[5] J. Akimitsu[3], and M. Braden[4]

[1]*National Institute of Advanced Industrial Science and Technology (AIST), Tsukuba, Ibaraki 305-8568, Japan.* [2] *Heinz Maier-Leibnitz Zentrum (MLZ), Technische Universität München, D-85748 Garching, Germany,* [3]*Aoyama Gakuin University, Sagamihara 252-5258, Japan.* [4]*University zu Köln, Germany,* [5]*Laboratoire Léon Brillouin (LLB), C.E.A./C.N.R.S., F-91191 Gif-sur-Yvette Cedex, France.* [*]*E-mail: c.lee@aist.go.jp*

1. Single crystal growth

Single crystals of $Ba_{1-x}K_xFe_2As_2$ were grown by the self-flux method [S1]. The starting materials were Ba (99.9%), K (99.9%), Fe (99.99%), and As (99.9999%). First, we synthesized the precursors KAs, BaAs, $Fe_2As$ and FeAs. Starting materials were put into alumina crucibles encapsulated into stainless steel containers filled with dried $N_2$. They were then heated for 10 h at 650 °C for KAs and BaAs and 900 °C for FeAs and $Fe_2As$. The obtained precursors were mixed and encapsulated into a stainless steel container with an alumina crucible again. The container was heated up to 900 °C and then cooled down to 650 °C at a rate of 1°C / h. The single crystals thus obtained had a tabular shape.

2. Results of inelastic neutron scattering

Figure S1 shows the Q-spectra of $Ba_{1-x}K_xFe_2As_2$ (x=0.50, 0.66, 0.77 and 0.84) at E < $E_{res}$ and E ~ $E_{res}$ above and below $T_c$. A remarkable enhancement of intensity was observed at E ~ $E_{res}$ upon cooling for x = 0.50. The enhancement weakened as doping increased and became small for x = 0.84. Suppression of intensity by cooling was observed at E < $E_{res}$. For x = 0.50, magnetic peaks at E = 5 meV vanished below $T_c$. On the other hand, weak incommensurate intensity at E = 3 or 4 meV remained for x ≥ 0.77, although the intensity was suppressed upon cooling.

Energy dependences of magnetic signals at T ~ $T_c$ and T < $T_c$ with L = 2 are shown in Figure S2. Figure S3 shows χ"(q,ω) at Q = (0.5 ± δ, 0.5 ± δ, L = *even*). The behavior of χ"(q,ω) at L = even is essentially the same as that at L = odd. The difference in $E_{res}$ between L = even and odd is within 1 meV.

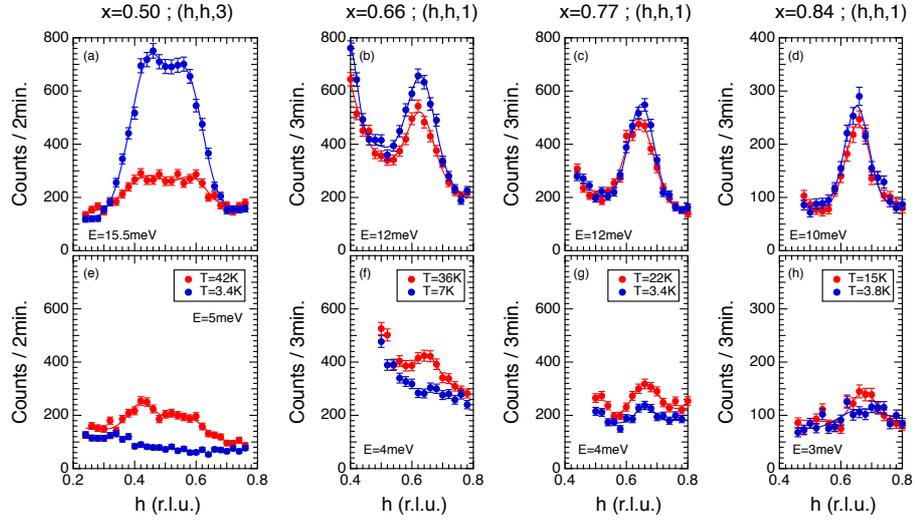

Figure S1. (a) Q-spectra of $Ba_{1-x}K_xFe_2As_2$ at $E \sim E_{res}$ and $E < E_{res}$ at $T \sim T_c$ and $T < T_c$ for (a) x = 0.50, (b) x = 0.66, (c) x = 0.77 and (d) x = 0.84. Solid lines indicate Gaussian fits.

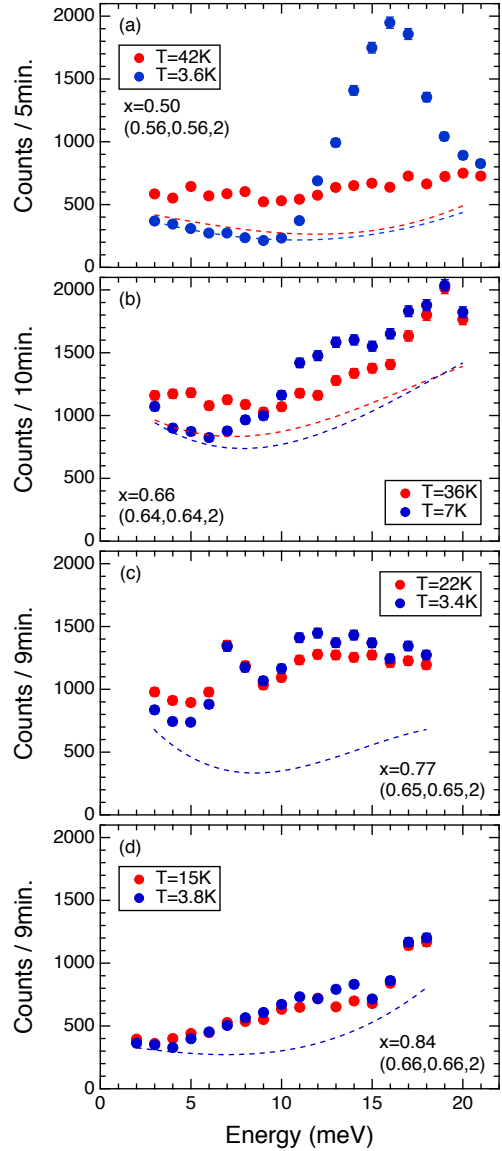

Figure S2. Energy dependences of magnetic signals at Q = (0.5 ± δ, 0.5 ± δ, 2) at T ∼ $T_c$ and T < $T_c$ for (a) x = 0.50, (b) x = 0.66, (c) x = 0.77 and (d) x = 0.84. Data for (d) were obtained at 2T1 and others were obtained at PUMA. Dashed lines describe the background at T ∼ $T_c$ (red) and T < $T_c$ (blue) determined by averaging the background at (a) (0.28,0.28,3) and (0.695,0.695,0); (b) (0.5,0.5,2), (0.5,0.5,3), (0.22,0.22,2) and (0.78,0.78,2); (c) (0.2,0.2,3.66) and (d) (0.2,0.2,3.74) and (0.5,0.5,2.95).

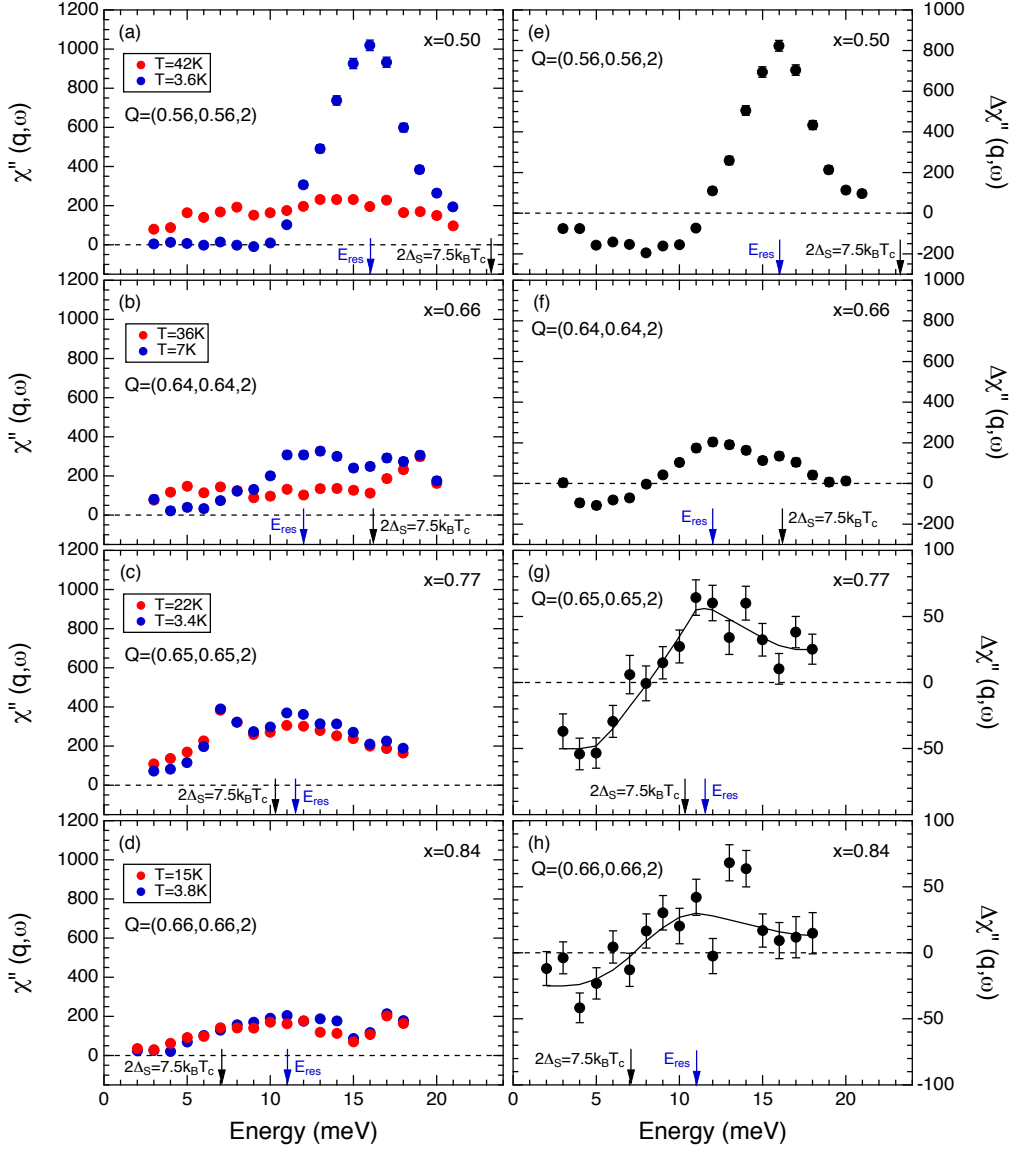

Figure S3. (a-d) Energy dependences of $\chi''(q,\omega)$ at $Q = (0.5 \pm \delta, 0.5 \pm \delta, 2)$ at $T \sim T_c$ and $T < T_c$ for (a) $x = 0.50$, (b) $x = 0.66$, (c) $x = 0.77$ and (d) $x = 0.84$. (e-h) Difference in $\chi''(q,\omega)$ between $T \sim T_c$ and $T < T_c$. The solid lines are a guide to the eye.